\documentclass[12pt]{article}
\usepackage{a4wide}
\usepackage{latexsym}
\usepackage{amsmath}
\usepackage{amsfonts}
\usepackage{amscd}
\usepackage{cite}
\usepackage{axodraw}

\usepackage{pslatex}
\usepackage{graphicx}
\usepackage[latin1]{inputenc}
\usepackage[T1]{fontenc}

\def\bq{\begin{eqnarray}}
\def\eq{\end{eqnarray}}

\def\eps{\varepsilon}

\begin{document}

\thispagestyle{empty}

\begin{flushright}
  MZ-TH/07-19 \\
% version of \today
\end{flushright}

\vspace{1.5cm}

\begin{center}
  {\Large\bf Periods and Feynman integrals\\
  }
  \vspace{1cm}
  {\large Christian Bogner ${}^{a,b}$ and Stefan Weinzierl ${}^{a}$\\
  \vspace{1cm}
      {\small ${}^{a}$ \em Institut f{\"u}r Physik, Universit{\"a}t Mainz,}\\
      {\small \em D - 55099 Mainz, Germany}\\
  \vspace{2mm}
      {\small ${}^{b}$ \em Department of Mathematical Sciences, University of Durham}\\
      {\small \em Durham DH1 3LE, United Kingdom}\\
  } 
\end{center}

\vspace{2cm}

% abstract ---------------------------------------
\begin{abstract}\noindent
  { 
   We consider multi-loop integrals in dimensional regularisation and the corresponding Laurent series.
   We study the integral in the Euclidean region and where all ratios of invariants and masses have rational values.
   We prove that in this case all coefficients of the Laurent series are periods.
  }
\end{abstract}

\vspace*{\fill}

% main text ------------------------------------
\newpage

\section{Introduction}
\label{sec:intro}

The calculation of loop integrals in perturbative quantum field theories
is essential for accurate and precise theoretical predictions to be compared
to experiments.
The calculation of loop integrals is complicated by the occurrence 
of ultraviolet and infrared singularities.
Ultraviolet divergences are related to the high-energy behaviour of the integrals.
Renormalisation absorbs these divergences into renormalisation constants.
Infrared divergences may occur if massless particles are present in the theory.
For infrared-safe observables they cancel in the final result, when summed over all
degenerate states.

Dimensional regularisation \cite{'tHooft:1972fi,Bollini:1972ui,Cicuta:1972jf}
is usually employed to regularise these singularities.
Within dimensional regularisation one considers the loop integral in $D$ space-time dimensions
instead of the usual four space-time dimensions.
The result is expanded as a Laurent series in the parameter $\eps=(4-D)/2$, describing the deviation
of the $D$-dimensional space from the usual four-dimensional space.
The singularities manifest themselves as poles in $1/\eps$.
Each loop can contribute a factor $1/\eps$ from the ultraviolet divergence and a factor $1/\eps^2$
from the infrared divergences. 
Therefore an integral corresponding to a graph with $l$ loops can have poles up to $1/\eps^{2l}$.

In this paper we consider multi-loop integrals in dimensional regularisation 
and the corresponding Laurent series.
It is an interesting question to ask, what type of numbers the coefficients of the Laurent series are.
These numbers certainly depend on the momenta of the external particles and the masses of the particles
propagating in the loops.
In order to keep the loop integral dimensionless, one introduces an additional arbitrary mass $\mu$ 
and the final result will depend on this mass as well.
For scalar loop integrals the dependence on the external momenta is only through Lorentz
invariants like $(p_1+p_2+...+p_j)^2$.
Let us assume that all ratios of these invariants and masses are rational numbers.
They are certainly real numbers. Since the rational numbers are dense in the real numbers, this is no severe
restriction.
Let us further assume that we consider the loop integral in the Euclidean region, meaning that
all masses are positive or zero and all invariants are negative or zero.
From explicit calculations we know, that for all one-loop integrals
the $\eps^0$-coefficient involves only rational numbers, logarithms and dilogarithms.
In two-loop integrals we encounter multiple polylogarithms, which
have been studied extensively in the literature
by physicists \cite{Remiddi:1999ew,Vermaseren:1998uu,Gehrmann:2000zt,Gehrmann:2001pz,Gehrmann:2001jv,Gehrmann:2002zr,Moch:2001zr,Weinzierl:2004bn,Vollinga:2004sn,Blumlein:1998if,Blumlein:2003gb,Korner:2005qz,Kalmykov:2006hu,Maitre:2007kp}
and mathematicians \cite{Borwein,Hain,Goncharov,Goncharov:2001,Goncharov:2002,Goncharov:2002b,Gangl:2000,Gangl:2002,Minh:2000,Cartier:2001,Ecalle,Racinet:2002,Brown:2004,Brown:2006,Brown:2008}.
For the massless two-loop two-point function it has been shown that all coefficients of the Laurent expansion
can be expressed in terms of multiple zeta values \cite{Bierenbaum:2003ud}.
As a last example let us mention that there is strong numerical evidence
that the three-loop two-point function with equal internal masses involves 
elliptic integrals \cite{Laporta:2002pg}.
We are therefore tempted to ask, what is the unifying theme for all these examples?
Is there a common set of numbers or functions, to which all these examples belong?
We observe that common to all examples is the fact, that these functions yield periods when evaluated
with rational numbers as arguments. Periods are special numbers and are introduced in
section~\ref{sect:periods}.
This raises the question if the observation we made for the few examples above
can be shown to hold for all loop integrals.

In this paper we show that this is indeed the case and prove a rather general result:
With the assumptions above (Euclidean region and all ratios of invariants and masses rational)
we show that all coefficients of the Laurent series of an arbitrary scalar multi-loop integral
are periods.
This result puts strong restrictions on the class of functions, which can appear in the calculation
of multi-loop integrals.
It is well known that a general multi-loop integral involving tensor structures in the numerator can
be reduced to scalar integrals. An algorithm for this reduction is reviewed in section~\ref{sect:tensorreduction}.
Therefore our result extends to all multi-loop integrals.

Our result extends and generalises a theorem
by Belkale and Brosnan \cite{Belkale:2003} on the Laurent expansion of Igusa local zeta functions.
Our proof is constructive and based on recent work on the resolution of singularities
for multi-loop integrals \cite{Bogner:2007cr}.
The proof uses the close relationship of Feynman integrals with algebraic geometry, in particular
the equivalence of iterated sector decomposition \cite{Hepp:1966eg,Roth:1996pd,Binoth:2000ps,Binoth:2003ak}
with Hironaka's polyhedra game \cite{Hironaka:1964,Spivakovsky:1983}.
Methods of algebraic geometry have been introduced in the context of Feynman integrals
in \cite{Bloch:2005}.

This paper is organised as follows:
In section~\ref{sect:periods} we introduce the set of periods.
Section~\ref{sect:feynman_integrals} reviews basic facts about scalar multi-loop integrals.
Section~\ref{sect:tensorreduction} shows that all loop integrals occurring in a quantum field theory can 
be reduced to these scalar integrals.
In section~\ref{sect:theorem} we state the main theorem of this paper and present a proof.
Section~\ref{sect:conclusions} contains the conclusions.
In the appendices we discuss Hironaka's polyhedra game and the Laurent expansion
of the prefactors accompanying the loop integrals.
A third appendix discusses the relation to the work of \cite{Bloch:2005} for the sequence of blow-ups.

% -----------------------------------------------------------------------------

\section{Periods}
\label{sect:periods}

Periods are special numbers.
Before we give the definition, let us start with some sets of numbers:
The natural numbers $\mathbb{N}$,
the integer numbers $\mathbb{Z}$,
the rational numbers $\mathbb{Q}$,
the real numbers $\mathbb{R}$ and 
the complex numbers $\mathbb{C}$
are all well-known. More refined is already the set of algebraic numbers, 
denoted by $\bar{\mathbb{Q}}$.
An algebraic number is a solution of a polynomial equation with rational
coefficients:
\bq
 x^n + a_{n-1} x^{n-1} + \cdots + a_1 x + a_0 & = & 0,
 \;\;\; a_j \in \mathbb{Q}.
\eq
As all such solutions lie in $\mathbb{C}$, the set of algebraic numbers $\bar{\mathbb{Q}}$ 
is a sub-set of
the complex numbers $\mathbb{C}$.
Numbers which are not algebraic are called transcendental.
The sets $\mathbb{N}$, $\mathbb{Z}$, $\mathbb{Q}$ and $\bar{\mathbb{Q}}$ are countable, whereas
the sets $\mathbb{R}$, $\mathbb{C}$ and the set of transcendental numbers are uncountable.

Periods are a countable set of numbers, lying between $\bar{\mathbb{Q}}$ and $\mathbb{C}$.
There are several equivalent definitions for periods.
Kontsevich and Zagier gave the following definition \cite{Kontsevich:2001}:
A period is a complex number whose real and imaginary parts are values
of absolutely convergent integrals of rational functions with rational coefficients,
over domains in $\mathbb{R}^n$ given by polynomial inequalities with rational coefficients.
Domains defined by polynomial inequalities with rational coefficients
are called semi-algebraic sets.

We denote the set of periods by $\mathbb{P}$. The algebraic numbers are contained in the set of periods:
$\bar{\mathbb{Q}} \in \mathbb{P}$.
In addition, $\mathbb{P}$ contains transcendental numbers, an example for such a number is $\pi$:
\bq
 \pi & = & \iint\limits_{x^2+y^2\le1} dx \; dy.
\eq
The integral on the r.h.s. clearly shows that $\pi$ is a period.
On the other hand, it is conjectured that the basis of the natural logarithm $e$
and Euler's constant $\gamma_E$
are not periods.

We need a few basic properties of periods:
The set of periods $\mathbb{P}$ is a $\bar{\mathbb{Q}}$-algebra \cite{Kontsevich:2001,Friedrich:2005}.
In particular the sum and the product of two periods are again periods.
This is immediate for the multiplication of two periods: Let $a$ and $b$ be two periods,
given through
\bq
 a = \int\limits_{G_1} d^{n}x \; f(x),
 & &
 b = \int\limits_{G_2} d^{m}y \; g(y),
\;\;\;
 G_1 \subset \mathbb{R}^{n},
 \;\;\;
 G_2 \subset \mathbb{R}^{m},
\eq
where $f(x)$ and $g(y)$ are rational functions with rational coefficients.
Then
\bq
 a \cdot b & = & \int\limits_{G_1 \times G_2} d^nx \; d^my \; \left[ f(x) g(y) \right].
\eq
For the sum of two periods we consider
\bq
 G = G_1 \times \{ 0 \} \times [0,1]^m
     \;\; \cup \;\;
     [0,1]^n \times \{ 1 \} \times G_2
     \;\; \subset \;\;
     \mathbb{R}^n \times \mathbb{R} \times \mathbb{R}^m.
\eq
Then
\bq
 a + b & = & \int\limits_{G} d^nx \; d^my \; \left[ (1-t) f(x) + t g(y) \right],
\eq
where $t$ is the coordinate of the middle factor $\mathbb{R}$ of 
$\mathbb{R}^n \times \mathbb{R} \times \mathbb{R}^m$.

The defining integrals of periods have integrands, which are rational
functions with rational coefficients.
For our purposes this is too restrictive, as we will encounter
logarithms as integrands as well.
The following lemma is easy to prove:

{\bf Lemma 1}: Let $G \subset \mathbb{R}^n$ be a semi-algebraic set and $f(x)$ and $g(x)$
two rational functions with rational coefficients.
Assume that the integral
\bq
 I & = & \int\limits_G d^nx \; f(x) \; \ln g(x) 
\eq
is absolutely convergent.
Then $I$ is a period.

Proof: We have
\bq
 \int\limits_G d^nx \; f(x) \; \ln g(x) 
 & = &
 \int\limits_G d^nx \; f(x) \; \int\limits_0^1 dt \; \frac{g(x)-1}{(g(x)-1) t + 1}
 \;\; = \;\;
 \int\limits_{G'} d^nx \; dt \; \frac{f(x)(g(x)-1)}{(g(x)-1) t + 1},
\eq
where $G' \subset \mathbb{R}^{n+1}$ and $(x_1,...,x_n,t) \in G'$ if $(x_1,...,x_n) \in G$ and
$t \in [0,1]$.
Clearly $G'$ is again a semi-algebraic set.
Therefore the integral $I$ is a period.

With the same technique of introducing additional variables one shows for rational functions
$f_1(x)$, $g_1(x)$, $f_2(x)$ and $g_2(x)$, all of them with rational coefficients, 
the following:
If the integral
\bq
 J & = & \int\limits_G d^nx \; 
  \left\{
  f_1(x) \; \ln g_1(x)  
  +
  f_2(x) \; \ln g_2(x) 
  \right\}
\eq
is absolutely convergent, then it is a period.
Here we have
\bq
 J & = & \int\limits_{G'} d^nx \; dt_1 \; dt_2 \;
 \left(
  \frac{f_1(x)(g_1(x)-1)}{(g_1(x)-1) t_1 + 1}
  +
  \frac{f_2(x)(g_2(x)-1)}{(g_2(x)-1) t_2 + 1}
 \right),
\eq
with $(x_1,...,x_n,t_1,t_2) \in G' \subset \mathbb{R}^{n+2}$ if
$(x_1,...,x_n) \in G$ and $t_1 \in [0,1]$ and $t_2 \in [0,1]$.
This shows that $G'$ is a semi-algebraic set and therefore $J$ is a period.

As a final example we consider with the definitions as above
the integral
\bq
 K & = & \int\limits_G d^nx \; f(x) \; \ln g_1(x) \; \ln g_2(x)
 \nonumber \\
 & = &
   \int\limits_{G} d^nx 
   \int\limits_0^1 dt_1
   \int\limits_0^1 dt_2
   \;
   f(x)
   \;
   \frac{g_1(x)-1}{(g_1(x)-1) t_1 + 1}
   \;
   \frac{g_2(x)-1}{(g_2(x)-1) t_2 + 1}.
\eq
If the integral $K$ is absolutely convergent, then it is a period.

Clearly these examples can be combined and iterated and we conclude that in the defining integral
for a period also integrands, which are linear combinations 
of products of rational functions with logarithms of rational functions,
all of them with rational coefficients, are allowed.

% -----------------------------------------------------------------------------

\section{Feynman integrals}
\label{sect:feynman_integrals}

In this section we introduce the central object of our investigations, Feynman loop integrals.
Most of the material in this section is well-known to physicists 
and can be found in many text-books \cite{Itzykson:1980rh,Smirnov:2004ym,Smirnov:2006ry,Weinzierl:2006qs}.
This section is included for mathematicians with little background on Feynman diagrams, but interested
in the most important facts.
In this section we restrict ourselves to scalar integrals.
These are integrals, where
the numerator of the integrand is independent of the loop momentum.
A priori more complicated cases, where the loop momentum appears in the numerator might occur.
However, there is a general reduction algorithm, which reduces these tensor integrals
to scalar integrals \cite{Tarasov:1996br,Tarasov:1997kx}.
This reduction algorithm is reviewed in section~\ref{sect:tensorreduction}.

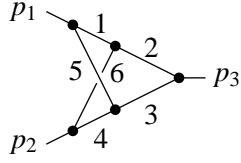
\begin{figure}
\begin{center}
\begin{picture}(57,40)(10,25)
\Line(50,30)(60,30)
\Vertex(50,30){2}
\Line(0,5)(50,30)
\Line(0,55)(50,30)
\Vertex(10,10){2}
\Vertex(10,50){2}
\Vertex(26,18){2}
\Vertex(26,42){2}
\Line(10,50)(26,18)
\Line(10,10)(19,28)
\Line(21,32)(26,42)
\Text(37,38)[bl]{\small$2$}
\Text(37,20)[tl]{\small$3$}
\Text(18,47)[bl]{\small$1$}
\Text(18,11)[tl]{\small$4$}
\Text(14,33)[r]{\small$5$}
\Text(24,32)[l]{\small$6$}
\Text(-3,55)[r]{\small $p_1$}
\Text(-3,5)[r]{\small $p_2$}
\Text(63,30)[l]{\small $p_3$}
\end{picture}
\end{center} 
\caption{\label{example_graph} An example of a two-loop Feynman graph with three external legs.}
\end{figure}

To set the scene let us consider a scalar Feynman graph $G$.
Fig.~\ref{example_graph} shows an example.
In this example there are three external lines and six internal lines. 
The momenta flowing in or out through the external lines
are labelled $p_1$, $p_2$ and $p_3$ and can be taken as fixed vectors.
They are constrained by momentum conservation: If all momenta are taken to flow outwards,
momentum conservation requires that
\bq
 p_1 + p_2 + p_3  & = & 0.
\eq
At each vertex of a graph we have again momentum conservation: The sum of all momenta flowing into the vertex
equals the sum of all momenta flowing out of the vertex.
A graph, where the external momenta determine uniquely all internal momenta is called a tree graph.
It can be shown that such a graph does not contain any closed circuit.

In contrast, graphs which do contain one or more closed circuits are called loop graphs.
If we have to specify besides the external momenta in addition 
$l$ internal momenta in order to determine uniquely all
internal momenta we say that the graph contains $l$ loops.
In this sense, a tree graph is a graph with zero loops and the graph in fig.~\ref{example_graph} contains two loops.
Let us agree that we label the $l$ additional internal momenta by $k_1$ to $k_l$.

Feynman rules allow us to translate a Feynman graph into a mathematical formula.
For a scalar graph we have to substitute for each internal line $j$ a propagator
\bq
 \frac{i}{q_j^2-m_j^2}.
\eq
Here, $q_j$ is the momentum flowing through line $j$. It is a linear combination of the external 
momenta $p$ and the loop momenta $k$:
\bq
 q_j & = & q_j(p,k).
\eq
$m_j$ is the mass of the particle of line $j$.
The Feynman rules tell us also to integrate for each loop over the loop momentum:
\bq
 \int \frac{d^4k_r}{(2\pi)^4}
\eq
However, there is a complication:
If we proceed naively and write down for each loop an integral over 
four-dimensional Minkowski space, we end up with ill-defined integrals, since these integrals
may contain ultraviolet or infrared divergences.
Therefore the first step is to make these integrals well-defined by introducing a regulator.
There are several possibilities how this can be done, but the
method of dimensional regularisation 
\cite{'tHooft:1972fi,Bollini:1972ui,Cicuta:1972jf}
has almost become a standard, as the calculations in this regularisation
scheme turn out to be the simplest.
Within dimensional regularisation one replaces the four-dimensional integral over the loop momentum by an
$D$-dimensional integral, where $D$ is now an additional parameter, which can be a non-integer or
even a complex number.
We consider the result of the integration as a function of $D$ and we are interested in the behaviour of this 
function as $D$ approaches $4$.
It is common practice to parameterise the
deviation of $D$ from $4$ by
\bq 
 D & = & 4 - 2\eps.
\eq
The divergences in loop integrals will manifest themselves in poles in $1/\eps$. 
In an $l$-loop integral ultraviolet divergences will lead to poles $1/\eps^l$ at the worst, whereas
infrared divergences can lead to poles up to $1/\eps^{2l}$.
We will also encounter integrals, where the dimension is shifted by units of two.
In these cases we often write
\bq
\label{shifted_dim}
 D & = & 2m - 2\eps,
\eq
where $m$ is an integer, and we are again interested in the Laurent series in $\eps$.

Let us now consider a generic scalar $l$-loop integral $I_G$ 
in $D=2m-2\eps$ dimensions with $n$ propagators,
corresponding to a graph $G$.
Let us further make a slight generalisation: 
For each internal line $j$ the corresponding propagator
in the integrand can be raised to a power $\nu_j$.
Therefore the integral will depend also on the numbers $\nu_1$,...,$\nu_n$.
In this paper it is sufficient to consider only the case, where all exponents are natural numbers: $\nu_j \in {\mathbb N}$.
We define the Feynman integral by
\bq
\label{eq0}
I_G  & = &
 \frac{\prod\limits_{j=1}^{n}\Gamma(\nu_j)}{\Gamma(\nu-lD/2)}
 \left( \mu^2 \right)^{\nu-l D/2}
 \int \prod\limits_{r=1}^{l} \frac{d^Dk_r}{i\pi^{\frac{D}{2}}}\;
 \prod\limits_{j=1}^{n} \frac{1}{(-q_j^2+m_j^2)^{\nu_j}},
\eq
with $\nu=\nu_1+...+\nu_n$.
The momenta $q_j$ of 
the propagators are linear combinations of the external momenta and the loop
momenta.
In eq.~(\ref{eq0}) there are some overall factors, which are inserted for convenience:
There is a prefactor consisting of Gamma-functions.
The arbitrary mass scale $\mu$ ensures that eq.~(\ref{eq0}) is dimensionless.
The integral measure is now $d^Dk/(i \pi^{D/2})$ instead of $d^Dk/(2 \pi)^D$, and each propagator
is multiplied by $i$.
The prefactors are chosen such that after Feynman parameterisation the Feynman integral
has a simple form.
The change of prefactors and the Laurent expansion of the neglected prefactors 
is discussed in detail in the appendix \ref{sect:prefactors}.

How to perform the $D$-dimensional loop integrals ?
The first step is to convert the products of propagators into a sum.
This can be done with the Feynman parameter technique.
In its full generality it is also applicable to cases, where each factor in the denominator is raised to 
some power $\nu_j$.
The formula reads:
\bq
\label{feynman_parameterisation}
 \prod\limits_{j=1}^{n} \frac{1}{P_{j}^{\nu_{j}}} 
 & = &
 \frac{\Gamma(\nu)}{\prod\limits_{j=1}^{n} \Gamma(\nu_{j})}
 \;\;
 \int\limits_{x_j \ge 0} d^nx 
 \;\;
 \delta(1-\sum\limits_{i=1}^{n} x_{i})
 \;
 \left( \prod\limits_{j=1}^{n} x_{j}^{\nu_{j}-1} \right)
 \;
 \left( \sum\limits_{i=1}^{n} x_{i} P_{i} \right)^{-\nu}.
\eq
Applied to eq.~(\ref{eq0}) we have
\bq
 \sum\limits_{i=1}^{n} x_{i} P_{i} & = & \sum\limits_{i=1}^{n} x_{i} (-q_i^2+m_i^2).
\eq
One can now use translational invariance of the $D$-dimensional loop integrals and shift each loop
momentum $k_r$ to complete the square, such that the integrand depends only on $k_r^2$.
Then all $D$-dimensional loop integrals can be performed.
The master formula for a single loop integration is:
\bq
\label{master_loop}
\int \frac{d^{D}k}{i \pi^{\frac{D}{2}}}
\frac{(-k^2)^a}{\left[ -U k^2 + F\right]^\nu} 
 & = &
 \frac{\Gamma(\frac{D}{2}+a)}{\Gamma(\frac{D}{2})}
 \frac{\Gamma(\nu-\frac{D}{2}-a)}{\Gamma(\nu)} 
 \frac{U^{-\frac{D}{2}-a}}{F^{\nu-\frac{D}{2}-a}}.
\eq
The functions $U$ and $F$ depend on the Feynman parameters 
and are obtained after Feynman parametrisation from completing the square.
In eq.~(\ref{master_loop}) we allowed additional powers $(-k^2)^a$ of the loop momentum in the numerator.
This is a slight generalisation and will be useful later.

With eq.~(\ref{master_loop}) one can perform all $D$-dimensional loop integrals iteratively.
As the integrals over the Feynman parameters still remain,
this allows us to treat the
$D$-dimensional loop integrals for Feynman parameter integrals.
One arrives at the following Feynman parameter integral,
which is the central object of investigation of this article:
\bq
\label{eq1}
I_G  & = &
 \left( \mu^2 \right)^{\nu-l D/2}
 \int\limits_{x_j \ge 0}  d^nx \;
 \delta(1-\sum_{i=1}^n x_i)\,
 \left( \prod\limits_{j=1}^n x_j^{\nu_j-1} \right)
 \frac{{\mathcal U}^{\nu-(l+1) D/2}}{{\mathcal F}^{\nu-l D/2}}.
\eq
The functions ${\mathcal U}$ and $\mathcal F$ depend on the Feynman parameters.
If one expresses
\bq
 \sum\limits_{j=1}^{n} x_{j} (-q_j^2+m_j^2)
 & = & 
 - \sum\limits_{r=1}^{l} \sum\limits_{s=1}^{l} k_r M_{rs} k_s + \sum\limits_{r=1}^{l} 2 k_r \cdot Q_r - J,
\eq
where $M$ is a $l \times l$ matrix with scalar entries and $Q$ is a $l$-vector
with four-vectors as entries,
one obtains
\bq
 {\mathcal U} = \mbox{det}(M),
 & &
 {\mathcal F} = \mbox{det}(M) \left( - J + Q M^{-1} Q \right).
\eq
Alternatively,
the functions ${\mathcal U}$ and ${\mathcal F}$ can be derived 
from the topology of the corresponding Feynman graph $G$.
Cutting $l$ lines of a given connected $l$-loop graph such that it becomes a connected
tree graph $T$ defines a chord ${\mathcal C}(T,G)$ as being the set of lines 
not belonging to this tree. The Feynman parameters associated with each chord 
define a monomial of degree $l$. The set of all such trees (or 1-trees) 
is denoted by ${\mathcal T}_1$.  The 1-trees $T \in {\mathcal T}_1$ define 
${\mathcal U}$ as being the sum over all monomials corresponding 
to the chords ${\mathcal C}(T,G)$.
Cutting one more line of a 1-tree leads to two disconnected trees $(T_1,T_2)$, or a 2-tree.
${\mathcal T}_2$ is the set of all such  pairs.
The corresponding chords define  monomials of degree $l+1$. Each 2-tree of a graph
corresponds to a cut defined by cutting the lines which connected the two now disconnected trees
in the original graph. 
The square of the sum of momenta through the cut lines 
of one of the two disconnected trees $T_1$ or $T_2$
defines a Lorentz invariant
\bq
s_{T} & = & \left( \sum\limits_{j\in {\mathcal C}(T,G)} p_j \right)^2.
\eq   
The function ${\mathcal F}_0$ is the sum over all such monomials times 
minus the corresponding invariant. The function ${\mathcal F}$ is then given by ${\mathcal F}_0$ plus an additional piece
involving the internal masses $m_j$.
In summary, the functions ${\mathcal U}$ and ${\mathcal F}$ are obtained from the graph as follows:
\bq
\label{eq0def}	
 {\mathcal U} 
 & = & 
 \sum\limits_{T\in {\mathcal T}_1} \Bigl[\prod\limits_{j\in {\mathcal C}(T,G)}x_j\Bigr]\;,
 \nonumber\\
 {\mathcal F}_0 
 & = & 
 \sum\limits_{(T_1,T_2)\in {\mathcal T}_2}\;\Bigl[ \prod\limits_{j\in {\mathcal C}(T_1,G)} x_j \Bigr]\, (-s_{T_1})\;,
 \nonumber\\
 {\mathcal F} 
 & = &  
 {\mathcal F}_0 + {\mathcal U} \sum\limits_{j=1}^{n} x_j m_j^2\;.
\eq
In general, ${\mathcal U}$ is a positive semi-definite function. 
Its vanishing is related to the  UV sub-divergences of the graph. 
Overall UV divergences, if present,
will always be contained in the  prefactor $\Gamma(\nu-l D/2)$. 
In the Euclidean region, ${\mathcal F}$ is also a positive semi-definite function 
of the Feynman parameters $x_j$.
The Euclidean region is defined as the region, where all invariants
$s_T$ are negative or zero.  
The vanishing of ${\mathcal F}$ is related to infrared divergences.
Note that this is only a necessary but not sufficient condition for the occurrence
of an infrared singularity.
If an infrared singularity occurs or not depends in addition on the external kinematics.

As an example for the functions ${\mathcal U}$ and ${\mathcal F}$
we consider the graph in fig.~\ref{example_graph}.
For simplicity we assume that all internal propagators are massless. Then the functions ${\mathcal U}$ and ${\mathcal F}$ read:
\bq
 {\mathcal U} & = & x_{15} x_{23} + x_{15} x_{46} + x_{23} x_{46},
 \nonumber \\
 {\mathcal F} & = & 
  \left( x_1 x_3 x_4 + x_5 x_2 x_6 + x_1 x_5 x_{2346} \right) \left( -p_1^2 \right) 
 \nonumber \\
 & &
  + \left( x_6 x_3 x_5 + x_4 x_1 x_2 + x_4 x_6 x_{1235} \right) \left( -p_2^2 \right) 
 \nonumber \\
 & &
  + \left( x_2 x_4 x_5 + x_3 x_1 x_6 + x_2 x_3 x_{1456} \right) \left( -p_3^2 \right).
\eq
Here we used the notation that $x_{ij...r} = x_i + x_j + ... + x_r$.

Before we close this section, let us consider one special case,
where the following three conditions are met:
\begin{enumerate}
\item The graph has no external lines or all invariants $s_T$ are zero.
\item All internal masses $m_j$ are equal to $\mu$.
\item All propagators occur with power $1$, i.e. $\nu_j=1$ for all $j$.
\end{enumerate}
In this case the Feynman parameter integral reduces to
\bq
\label{localigusazetafct}
I_G  & = &
 \int\limits_{x_j \ge 0}  d^nx \;
 \delta(1-\sum_{i=1}^n x_i)\,
 {\mathcal U}^{- D/2}.
\eq
This integral is a Igusa local zeta function (when viewed as a function of $D/2$)
and has been studied by Belkale and Brosnan in \cite{Belkale:2003}.

% -----------------------------------------------------------------------------

\section{Reduction to scalar integrals}
\label{sect:tensorreduction}

In the previous section we discussed scalar loop integrals, where
the numerator of the integrand is independent of the loop momentum.
This is not the most general case.
A priori more complicated cases, where the loop momentum appears in the numerator might occur.
However, there is a general reduction algorithm, which reduces these tensor integrals
to scalar integrals \cite{Tarasov:1996br,Tarasov:1997kx}.
The price we have to pay is that these scalar integrals involve higher powers of the propagators
and/or have shifted dimensions.
Therefore we considered in eq.~(\ref{shifted_dim}) shifted dimensions
and in eq.~(\ref{eq0}) arbitrary powers of the propagators.
As a consequence, integrals of the form as in eq.~(\ref{eq0}) 
are the most general loop integrals we have to consider.
In this section we review the basic features of the reduction algorithm.

We recall that the $D$-dimensional loop integrals over $k_j$ can be done loop by loop
by Feynman parameterisation, completing the square in the denominator and a shift in the
loop integration variables.
The presence of the loop momentum in the numerator has the following effect:
After the shift of the loop momentum, the (shifted) loop momentum as well as the 
Feynman parameter appear in the numerator.
Let us first discuss the presence of the shifted (loop) momentum.
We have to consider integrals of the form
\bq
\int \frac{d^{D }k}{i\pi^{D /2}} \; k^{\mu_1} k^{\mu_2} .... k^{\mu_r} \; f(k^2),
\eq
where $f(k^2)$ is a function depending only on $k^2$, but not on the individual components $k^\mu$.
Integrals with an odd number of the loop momentum in the numerator vanish by symmetry, while
integrals with an even number of the loop momentum can be related by Lorentz
invariance to the following integrals:
\bq
\int \frac{d^{D }k}{i\pi^{D /2}} k^\mu k^\nu f(k^2) & = & 
 - \frac{1}{D } g^{\mu\nu} \int \frac{d^{D }k}{i\pi^{D /2}} (-k^2) f(k^2), \\
\int \frac{d^{D }k}{i\pi^{D /2}} k^\mu k^\nu k^\rho k^\sigma f(k^2) & = & 
 \frac{1}{D (D +2)} 
  \left( g^{\mu\nu} g^{\rho\sigma} + g^{\mu\rho} g^{\nu\sigma} + g^{\mu\sigma} g^{\nu\rho} \right) 
  \int \frac{d^{D }k}{i\pi^{D /2}} (-k^2)^2 f(k^2). \nonumber
\eq
The generalisation to arbitrary higher tensor structures is obvious.
Recalling eq.~(\ref{master_loop}) we observe
that the dependency of the result on additional factors $(-k^2)^a$ occurs only in the combination 
$D/2+a$, apart from a trivial factor $\Gamma(D/2+a)/\Gamma(D/2)$. 
Therefore adding a power of $(-k^2)$ to the numerator is equivalent to consider the integral
without this power in dimensions $D+2$.

In addition, shifting the loop momentum like in $k'=k-x p$ introduces for tensor integrals 
the Feynman parameters $x_j$ in the numerator.
From the formula~(\ref{feynman_parameterisation})
we observe that 
a Feynman parameter $x$ in the numerator is equivalent to raising the power of the original
propagator by one unit: $\nu \rightarrow \nu+1$.
Therefore we can relate an integral, where a Feynman parameter occurs in the numerator
to a scalar integral, where the corresponding propagator is raised to a higher power.

In summary, we can express
all tensor integrals in terms of scalar integrals.
The price we have to pay is that these scalar integrals involve higher powers of the propagators
and/or have shifted dimensions.
Therefore integrals of the form as in eq.~(\ref{eq1}) are the most general integrals which we have to consider.

% -----------------------------------------------------------------------------

\section{The main theorem}
\label{sect:theorem}

Let us consider a general scalar multi-loop integral as in eq.~(\ref{eq1})
\bq
I_G  & = &
 \left( \mu^2 \right)^{\nu-l D/2}
 \int\limits_{x_j \ge 0}  d^nx \;
 \delta(1-\sum_{i=1}^n x_i)\,
 \left( \prod\limits_{j=1}^n x_j^{\nu_j-1} \right)
 \frac{{\mathcal U}^{\nu-(l+1) D/2}}{{\mathcal F}^{\nu-l D/2}}.
\eq
Let $m$ be an integer and set $D=2 m - 2 \eps$. Then this integral has 
a Laurent series expansion in $\eps$
\bq
 I_G & = & \sum\limits_{j=-2l}^\infty c_j \eps^j.
\eq
{\bf Theorem 1}: In the case where
\begin{enumerate}
\item all kinematical invariants $s_T$ are negative or zero, 
\item all masses $m_i$ and $\mu$ are positive or zero ($\mu\neq0$),
\item all ratios of invariants and masses are rational,
\end{enumerate}
the coefficients $c_j$ of the Laurent expansion are periods.

Conditions (1) and (2) define the Euclidean region, where the integral does not possess
an imaginary part.
Condition (3) restricts all ratios to rational numbers.
Since the rational numbers are dense in the real numbers, this is no severe
restriction.

Before we prove this theorem, we comment on two special cases.
We first discuss the coefficient $c_0$ in the case where the integral is convergent 
(and all propagators occur with power $1$).
In this case it is obvious, that $c_0$ is a period.
An example is given by the one-loop three-point function with three external masses
and vanishing internal masses. This integral is finite and no regularisation is needed.
Therefore the Laurent series starts at $\eps^0$. The coefficient $c_0$ is given by
\bq
c_0
  & = &
 \int\limits_{x_j \ge 0}  d^3x \;
 \frac{\delta(1-x_1-x_2-x_3)}{a_1 x_2 x_3 + a_2 x_3 x_1 + a_3 x_1 x_2},
 \nonumber \\
 & &
 a_1 = \frac{-(p_2+p_3)^2}{\mu^2},
 \;\;\;
 a_2 = \frac{-(p_3+p_1)^2}{\mu^2},
 \;\;\;
 a_3 = \frac{-(p_1+p_2)^2}{\mu^2}.
\eq
By assumption, the parameters $a_1$, $a_2$ and $a_3$ are positive rational numbers.
The integrand $1/(a_1 x_2 x_3 + a_2 x_3 x_1 + a_3 x_1 x_2)$ 
is clearly a rational function, integrated over a semi-algebraic set.
This shows that $c_0$ is a period.

The second case is given by integrals, where the conditions discussed at the end of section~\ref{sect:feynman_integrals}
are met:
The graph has no external lines or all invariants $s_T$ are zero,
all internal masses $m_j$ are equal to $\mu$ and 
all propagators occur with power $1$.
In this case the integral reduces to a Igusa local zeta function of the form
\bq
I_G  & = &
 \int\limits_{x_j \ge 0}  d^nx \;
 \delta(1-\sum_{i=1}^n x_i)\,
 {\mathcal U}^{- D/2}
\eq
and it has been shown by
Belkale and Brosnan \cite{Belkale:2003} that the coefficients of the Laurent expansion are periods.

We will actually prove a stronger version of theorem 1.
Consider the following integral
\bq
\label{basic_integral}
J & = &
 \int\limits_{x_j \ge 0} d^nx \;\delta(1-\sum_{i=1}^n x_i)
 \left( \prod\limits_{i=1}^n x_i^{a_i+\eps b_i} \right)
 \prod\limits_{j=1}^r \left[ P_j(x) \right]^{d_j+\eps f_j}.
\eq
The integration is over the standard simplex.
The $a$'s, $b$'s, $d$'s and $f$'s are integers.
The $P$'s are polynomials in the variables $x_1$, ..., $x_n$ with rational coefficients.
The polynomials are required to be non-zero
inside the integration region, but
may vanish on the boundaries of the integration region.
To fix the sign, let us agree that all polynomials are positive inside the integration region.
The integral $J$ has a Laurent expansion
\bq
 J & = & \sum\limits_{j=j_0}^\infty c_j \eps^j.
\eq
{\bf Theorem 2}: The coefficients $c_j$ of the Laurent expansion of the integral $J$ are periods.
\\
\\
Theorem 1 follows then from theorem 2 as the special case
$a_i=\nu_i-1$, $b_i=0$, $r=2$, $P_1={\cal U}$, $P_2={\cal F}$, 
$d_1+\eps f_1 = \nu-(l+1)D/2$ and $d_2+\eps f_2 = l D/2 - \nu$.
\\
\\
Proof of theorem 2:
To prove the theorem we will give an algorithm which expresses each coefficient $c_j$
as a sum of absolutely convergent integrals over the unit hypercube with integrands,
which are linear combinations 
of products of rational functions with logarithms of rational functions,
all of them with rational coefficients.
Let us denote this set of functions to which the integrands belong by ${\cal M}$:
Linear combinations of products of rational functions with logarithms of rational functions,
all of them with rational coefficients.

The unit hypercube is clearly a semi-algebraic set.
From section~\ref{sect:periods} we know, that absolutely convergent integrals
over semi-algebraic sets with integrands from the set ${\cal M}$ are periods.
In addition, the sum of periods is again a period.
Therefore it is sufficient to express each coefficient $c_j$ as a finite sum
of absolutely convergent integrals over the unit hypercube with integrands from ${\cal M}$.
To do so, we use iterated sector 
decomposition \cite{Hepp:1966eg,Roth:1996pd,Binoth:2000ps,Binoth:2003ak,Bogner:2007cr}.
The algorithm is described in detail in \cite{Bogner:2007cr}.
We proceed through the following steps:
\\
\\
Step 0: Starting from the original integral $J$ in eq.~(\ref{basic_integral})
we first convert all polynomials to homogeneous polynomials.
Due to the presence of the delta-function we have
\bq
 1 & = & x_1 + x_2 + ... + x_n.
\eq
and we can multiply each term in each polynomial $P_j$
by an appropriate power of $x_1 + x_2 + ... + x_n$.
\\
\\
Step 1: We can now assume that all polynomials are homogeneous.
We then decompose the integral into $n$ primary sectors as in
\bq
 \int\limits_{x_j \ge 0} d^nx & = &
 \sum\limits_{l=1}^n \int\limits_{x_j \ge 0} d^nx 
     \prod\limits_{i=1, i\neq l}^n \theta(x_l \ge x_i).
\eq
In the $l$-th primary sector we make the substitution
\bq
 x_j & = & x_l x_j' \;\;\;\mbox{for} \; j \neq l,
\eq
and integrate out the variable $x_l$ with the help of the delta-function.
Each primary sector is now a $(n-1)$-dimensional integral over the unit hypercube.
Note that in the general case this decomposition introduces an additional polynomial factor
\bq
 \left( 1 + \sum\limits_{j=1, j\neq l}^n x_j \right)^c,
&&
 c = -n - \sum\limits_{i=1}^n \left( a_i+\eps b_i \right) - \sum\limits_{j=1}^r h_j \left( c_j+\eps d_j\right),
\eq
where $h_j$ is the degree of the homogeneous polynomial $P_j$.
After this step and after a relabelling 
we deal with integrals of the form
\bq
\label{primary_integral}
 \int\limits_{0}^{1} d^{n}x
  \prod\limits_{i=1}^{n}x_{i}^{a_{i}+\epsilon b_{i}}
  \prod\limits_{j=1}^{r} \left[P_{j}(x)\right]^{d_{j}+\epsilon f_{j}},
\eq
where the integral is now over the unit hypercube
and the polynomials are positive semi-definite functions on the unit hypercube.
Zeros may only occur on coordinate subspaces.
Note that in general the polynomials $P_j$ are no longer homogeneous.
\\
\\
Step 2: We decompose the primary sectors iteratively into sub-sectors until each of the polynomials 
is of the form
\bq
\label{monomialised}
 P & = & C x_1^{m_1} ... x_n^{m_n} \left( 1 + P'(x) \right),
\eq
where $P'(x)$ is a polynomial in the variables $x_j$ not containing a constant term and $C$ is a rational number.
The term $1+P'(x)$ is now a positive definite function on the unit hypercube.
If $P$ is of the form~(\ref{monomialised}), we say that $P$ is monomialised.
In this case the monomial prefactor $x_1^{m_1} ... x_n^{m_n}$ can be factored out
and the remainder contains a constant term.
To convert $P$ into the form~(\ref{monomialised}) we choose a subset
$S=\left\{ \alpha_{1},\,...,\, \alpha_{k}\right\} \subseteq \left\{ 1, \,...\, n \right\}$
according to a strategy.
We decompose the $k$-dimensional hypercube into $k$ sub-sectors according to
\bq
 \int\limits_{0}^{1} d^{n}x & = & 
 \sum\limits_{l=1}^{k} 
 \int\limits_{0}^{1} d^{n}x
   \prod\limits_{i=1, i\neq l}^{k}
   \theta\left(x_{\alpha_{l}}\geq x_{\alpha_{i}}\right).
\eq
In the $l$-th sub-sector we make for each element of $S$ the
substitution
\bq
\label{substitution}
x_{\alpha_{i}} & = & x_{\alpha_{l}} x_{\alpha_{i}}' \;\;\;\mbox{for}\; i\neq l.
\eq
This procedure is iterated, until all polynomials are of the form~(\ref{monomialised}).
It is important to show that this can always be achieved in a finite number of iterations.
In ref.~\cite{Bogner:2007cr} we gave a strategy for choosing the subset $S$,
such that all polynomials are monomialised in a finite number of steps.
This was done by relating the problem of monomialising the polynomials 
to the problem of the resolution of singularities of an algebraic variety over a field of characteristic zero.
The monomialisation of a polynomial is equivalent to the special case, where the algebraic
variety is defined through a single polynomial.
For this special case Hironaka \cite{Hironaka:1964} invented his polyhedra game to illustrate the challenge
to find a constructive proof.
Spivakovsky \cite{Spivakovsky:1983} was the first to give a winning strategy for Hironaka's
polyhedra game and this winning strategy can be used in our case to ensure that the polynomials
are monomialised in a finite number of steps.
We review the relationship between Hironaka's polyhedra game 
and the problem of monomialising a polynomial in appendix \ref{sect:polyhedra_game}.

At the end of step 2 we have a finite number of sub-sector integrals.
Each sub-sector integral is of the form as in eq.~(\ref{primary_integral}),
where every $P_{j}$ is now different from zero in the whole integration
domain (including the boundaries). 
Hence the singular behaviour of the integral depends entirely on the factor
\bq
  \prod\limits_{i=1}^{n}x_{i}^{a_{i}+\epsilon b_{i}}.
\eq
Step 3: For every $x_{j}$ with $a_{j}<0$ we perform a Taylor
expansion around $x_{j}=0$ in order to extract the possible $\epsilon$-poles.
If we consider for the moment only one parameter $x_{j}$ we 
can write the
corresponding integral as 
\bq
 \int\limits_{0}^{1} dx_{j} \; x_{j}^{a_{j}+b_{j}\eps} \mathcal{I}(x_{j})
 = 
 \int\limits_{0}^{1} dx_{j} \; x_{j}^{a_{j}+b_{j}\eps}
   \left(\sum\limits_{p=0}^{\left|a_{j}\right|-1} \frac{x_{j}^{p}}{p!} \mathcal{I}^{(p)} 
         + \mathcal{I}^{(R)}(x_j)
   \right)
\eq
where we defined 
$\mathcal{I}^{(p)} = \left. \partial/\partial x_{j}^{p} \mathcal{I}(x_{j})\right|_{x_{j}=0}$.
The remainder term
\bq
 \mathcal{I}^{(R)}(x_j) & = & 
   \mathcal{I}(x_{j}) - \sum\limits_{p=0}^{\left|a_{j}\right|-1} \frac{x_{j}^{p}}{p!} \mathcal{I}^{(p)}
\eq
is by construction integrable in $x_j$ and 
does not lead to $\eps$-poles from the $x_{j}$-integration.
The integration in the pole part can be carried out analytically:
\bq
\label{analytic_integration}
 \int\limits_{0}^{1} dx_{j} \; x_{j}^{a_{j}+b_{j}\eps}
   \; \frac{x_{j}^{p}}{p!} \mathcal{I}^{(p)} 
  & = &
   \frac{1}{a_{j}+b_{j}\eps+p+1} \frac{\mathcal{I}^{(p)}}{p!}.
\eq
This procedure is repeated for all variables $x_j$ for which $a_j<0$.
At the end of step 3 we obtain a finite sum of integrals of the form
\bq
\label{finite_integral}
 K(\eps) & = & \frac{1}{g(\eps)} \int\limits_0^1 d^nx \; F(x,\eps),
\eq
with
\bq
 F(x,\eps) = \sum\limits_{j=1}^N f_j(x,\eps),
 & &
 f_j(x,\eps) = g_j(\eps) 
          \prod\limits_{i=1}^{n}x_{i}^{a^j_{i}+\epsilon b_{i}}
          \prod\limits_{k=1}^{r} \left[P^j_{k}(x)\right]^{d^j_{k}+\epsilon f_{k}}.
\eq
Here, $g(\eps)$ and $g_j(\eps)$ are polynomials in $\eps$ with integer coefficients.
$g(\eps)$ contains the explicit poles in $\eps$ from the integration in eq.~(\ref{analytic_integration}).
$P^j_k(x)$ is a polynomial with rational coefficients, non-vanishing on the unit hypercube.
Further we have $a^j_i, b_i, d^j_i, f_i \in \mathbb{Z}$.
The integral in eq.~(\ref{finite_integral}) 
\bq 
 \int\limits_0^1 d^nx \; F(x,\eps)
\eq
is convergent by construction
for all $\eps$ in a neighbourhood of $\eps=0$.
In one variable this integral is of the form
\bq
\label{int_in_one_variable}
 \int\limits_0^1 dx \; x^{\eps b} R(x,\eps),
\eq
where the function $R(x,\eps)$ does not contain any singularities on the integration domain and is therefore
bounded.
Therefore the integral in eq.~(\ref{int_in_one_variable}) is absolutely convergent for all $\eps$ with $|\eps| < |1/b|$.
\\
\\
Step 4: It remains to expand $K(\eps)$, $1/g(\eps)$ and $F(x,\eps)$ in $\eps$:
\bq
 K(\eps) = \sum\limits_{r=A}^\infty K_r \eps^r,
 \;\;\;
 \frac{1}{g(\eps)} = \sum\limits_{r=A}^\infty g_r \eps^r,
 \;\;\;
 F(x,\eps) = \sum\limits_{r=0}^\infty F_r(x) \eps^r,
 \;\;\;
 K_r = \sum\limits_{s=A}^r g_s \int\limits_0^1 d^nx \; F_{r-s}(x).
\eq
The expansion of the functions $1/g(\eps)$ and $g_j(\eps)$ yields rational numbers, for the other terms
we have
\bq
 x^{a+b\eps} & = & x^a \sum\limits_{k=0}^\infty \frac{b^k }{k!} \left( \ln x \right)^k \eps^k,
 \nonumber \\
 \left[ P(x) \right]^{d+\eps f} & = & 
   \left[ P(x) \right]^{d} \sum\limits_{k=0}^\infty \frac{f^k}{k!} \left( \ln\left(P(x)\right)\right)^k \eps^k.
\eq
Therefore the integrand $F_{r-s}(x)$ belongs to the set ${\cal M}$.

The integrals over $F_{r}(x)$ are absolutely convergent: In each variable we have integrals of the form
\bq
\label{finite_integral_2}
 \int\limits_0^1 dx \; \left( \ln x \right)^k R_r(x),
 \;\;\;\;
 k \in {\mathbb N}_0,
\eq
where the function $R_r(x)$ does not contain any singularities on the integration domain and is therefore bounded.
Therefore eq.~(\ref{finite_integral_2}) defines an absolutely convergent integral.
This completes the proof.

% -----------------------------------------------------------------------------

\section{Conclusions}
\label{sect:conclusions}

In this article we considered the Laurent expansion of multi-loop Feynman integrals in 
dimensional regularisation.
We studied the integral in the Euclidean region and where all ratios of invariants and masses 
have rational values.
We showed that in this case all coefficients of the Laurent series are periods (theorem 1).
This is an important result for the theory of Feynman integrals, as it restricts the class
of functions which can appear in the computation of Feynman integrals.
The proof of theorem 1 follows from a stronger result, stated in theorem 2:
Integrals of the form as in eq.~(\ref{basic_integral}) have a Laurent expansion, where all
coefficients are periods.
The proof of theorem 2 is constructive and based on iterated sector decomposition.

% -----------------------------------------------------------------------------

\begin{appendix}

\section{Hironaka's polyhedra game}
\label{sect:polyhedra_game}

In this appendix we show the equivalence of Hironaka's polyhedra game with the 
problem of monomialising a polynomial.

Hironaka's polyhedra game is played by two players, A and B. They are
given a finite set $M$ of points $m=\left(m_{1},\,...,\,m_{n}\right)$
in $\mathbb{N}_{+}^{n}$, the first quadrant of $\mathbb{N}^{n}$.
We denote by $\Delta \subset\mathbb{R}_{+}^{n}$ the positive convex hull of the set $M$.
It is given by the convex hull of the set
\bq
\bigcup\limits_{m\in M}\left(m+\mathbb{R}_{+}^{n}\right).
\eq
The two players compete in the following game:
\begin{enumerate}
\item Player A chooses a non-empty subset $S\subseteq\left\{ 1,\,...,\, n\right\}$.
\item Player B chooses one element $i$ out of this subset $S$. 
\end{enumerate}
Then, according
to the choices of the players, the components of all $\left(m_{1},\,...,\,m_{n}\right)\in M$
are replaced by new points $\left(m_{1}^{\prime},\,...,\,m_{n}^{\prime}\right)$,
given by:
\bq
\label{update_polyhedron}
m_{j}^{\prime} & = & m_{j}, \;\;\; \textrm{if }j\neq i, \nonumber \\
m_{i}^{\prime} & = & \sum_{j\in S} m_{j}-c,
\eq
where for the moment we set $c=1$. 
This defines the set $M^\prime$.
One then sets $M=M^\prime$ and goes back to step 1.
Player A wins the game if, after a finite number of moves, 
the polyhedron $\Delta$ is of the form
\bq
\label{termination}
 \Delta & = & m+\mathbb{R}_{+}^{n},
\eq
i.e. generated by one point.
If this never occurs, player $B$ has won.
The challenge of the polyhedra game is to show that player $A$ always has
a winning strategy, no matter how player $B$ chooses his moves.

Let us discuss the relation of Hironaka's polyhedra game to the sector decomposition of multi-loop
integrals.
Without loss of generality we can assume that we have just one polynomial $P$
in eq.~(\ref{basic_integral}).
(If there are several polynomials, we obtain a single polynomial by multiplying them together.
As only the zero-sets of the polynomials are relevant, the exponents can be neglected.)
The polynomial $P$ has the form
\bq
 P & = & 
  \sum\limits_{i=1}^p c_i x_1^{m_1^{(i)}} x_2^{m_2^{(i)}} ... x_n^{m_n^{(i)}}.
\eq
The $n$-tuple $m^{(i)}=\left(m^{(i)}_{1},\,...,\,m^{(i)}_{n}\right)$ defines a point in
$\mathbb{N}_{+}^{n}$ and 
$M=\left\{m^{(1)},\,...\,m^{(p)} \right\}$ is the set of all such points.
Substituting the
parameters $x_{j}$ according to equation~(\ref{substitution}) and factoring out
a term $x_{i}^c$ yields the same polynomial as replacing the powers
$m_{j}$ according to equation~(\ref{update_polyhedron}). 
In this sense, one iteration of the sector decomposition corresponds to one move in Hironaka's game. 
Reducing $P$ to the form~(\ref{monomialised})
is equivalent to achieving~(\ref{termination}) in the polyhedra game.
Finding a strategy which guarantees termination of the iterated sector decomposition 
corresponds to a winning strategy for player $A$ in the polyhedra game.
Note that we really need a strategy that guarantees player A's victory for every
choice player B can take, because the sector decomposition has to be carried out
in every appearing sector. In other words, we sample over all possible decisions of B.

% -----------------------------------------------------------------------------

\section{The prefactors}
\label{sect:prefactors}

In this appendix we discuss the prefactors, which we dropped in eq.~(\ref{eq0}).
From the Feynman rules we obtain for a loop integral
\bq
\tilde{I}_G  & = &
 g^{2 l}
 \left( \frac{e^{\gamma_E}}{4 \pi} \right)^{l \eps}
 \left( \mu^2 \right)^{\nu-l D/2}
 \int \prod\limits_{r=1}^{l} \frac{d^Dk_r}{(2\pi)^D} \;
 \prod\limits_{j=1}^{n} \frac{i}{(q_j^2-m_j^2)^{\nu_j}},
\eq
with $\nu=\nu_1+...+\nu_n$.
Here we included for each loop 
an additional coupling factor $g^2$ relative to a corresponding tree graph and
a factor $(e^{\gamma_E}/4/\pi)^\eps$ related to the
$\overline{MS}$-scheme \cite{Bardeen:1978yd}.
We have
\bq 
\tilde{I}_G & = & C_G \; I_G,
\eq
where $I_G$ is defined by eq.~(\ref{eq0}) and 
the prefactor $C_G$ is given by
\bq
 C_G & = & 
           g^{2 l}
           (-1)^\nu i^{n+l} (4\pi)^{-l D/2}
           \left( \frac{e^{\gamma_E}}{4 \pi} \right)^{l \eps}
           \frac{\Gamma(\nu-lD/2)}{\prod\limits_{j=1}^{n}\Gamma(\nu_j)}.
\eq
With $D=2m-2\eps$ we obtain
\bq
 C_G & = & \frac{(-1)^\nu i^{n+l} }{\prod\limits_{j=1}^{n}\Gamma(\nu_j)} 
           \left( e^{l \gamma_E \eps} \Gamma(\nu-lm+l\eps) \right) 
           \left( \frac{g^{2}}{(4\pi)^{m}} \right)^l.
\eq
Since all $\nu_j \in {\mathbb N}$, 
the term $(-1)^\nu i^{n+l}/\prod \Gamma(\nu_j)$ is obviously a period.
We can further show that the second term has a Laurent expansion in $\eps$, where all
coefficients are periods:
\bq
 e^{l \gamma_E \eps} \Gamma(\nu-lm+l\eps)
 & = &
 e^{l \gamma_E \eps} \Gamma(1+l\eps) \cdot \frac{\Gamma(\nu-lm+l\eps)}{\Gamma(1+l\eps)}.
\eq
Using the functional equation $\Gamma(x+1)=x\Gamma(x)$ one shows that
$\Gamma(\nu-lm+l\eps)/\Gamma(1+l\eps)$ is a rational function in $\eps$ with rational
coefficients.
The expansion of $\Gamma(1+\eps)$ is given by
\bq
\Gamma(1+\eps)  & = & 
  \exp \left( - \gamma_E \eps + \sum\limits_{n=2}^\infty \frac{(-1)^n}{n} \zeta_n \eps^n \right),
\eq
and therefore
\bq
 e^{l \gamma_E \eps} \Gamma(1+l\eps) & = &
  \exp \left( \sum\limits_{n=2}^\infty \frac{(-l)^n}{n} \zeta_n \eps^n \right).
\eq
The zeta values $\zeta_n$ are periods, therefore the product $e^{l \gamma_E \eps} \Gamma(\nu-lm+l\eps)$
has a Laurent expansion in $\eps$, where all coefficients are periods.
Note that Euler's constant $\gamma_E$, which is conjectured not to be a period, drops out in this 
expression.

This leaves the factor $g^{2 l}/(4\pi)^{l m}$. It is not known whether $1/\pi$ is a period or not.
Kontsevich and Zagier \cite{Kontsevich:2001} consider therefore the extended period ring
$\hat{\mathbb P} = {\mathbb P}[1/\pi]$, obtained by adding $1/\pi$ to the ring of periods.
In any case it is common practice to quote results of a perturbative calculation as an expansion in
$g^2/(4 \pi)^m$: For $D=4-2\eps$ and with $\alpha=g^2/(4\pi)$ the result is often expressed as
\bq
 \sigma & = & \sigma_0 \left[ 1 + \frac{\alpha}{4\pi} C_1 
                       + \left( \frac{\alpha}{4\pi} \right)^2 C_2 + ... \right]
\eq
In the coefficients $C_j$ the factors $g^2/(4\pi)^2=\alpha/(4\pi)$ have been explicitly factored out.
The results of this paper apply to the coefficients $C_j$.

\section{Other algorithms for the sequence of blow-ups}
\label{sect:BEK}

Triggered by the response of a referee we show in this appendix
that the algorithm for the construction of the sequence of blow-ups
given by Bloch, Esnault and Kreimer in \cite{Bloch:2005} does in general not
monomialize a polynomial.
We do this by giving an explicit counter-example.
Consider the polynomial
\bq
\label{counterexample}
 P & = & x_2^2 + x_1 x_3
\eq
with homogeneous coordinates $x_1$, $x_2$, $x_3$, $x_4$ on ${\mathbb P}^3$.
We will examine the polynomial $P$ in a neighbourhood of $(x_1,x_2,x_3,x_4)=(0,0,0,1)$.
Let us denote coordinate sub-spaces by
$L_{ij}: x_i=x_j=0$ and $L_{ijk}: x_i=x_j=x_k=0$.
The algorithm of \cite{Bloch:2005} determines first the set $S$ of coordinate
subspaces, on which $P$ vanishes and which are maximal.
In our case
\bq
 S & = & \{ L_{12}, L_{23} \}.
\eq
One then forms all possible intersections of elements of $S$, 
which defines the set
${\mathcal F}$:
\bq
 {\mathcal F} & = & \{ L_{12}, L_{23}, L_{123} \}.
\eq
${\mathcal F}_{min}$ contains the minimal elements of ${\mathcal F}$:
\bq
 {\mathcal F}_{min} & = & \{ L_{123} \}.
\eq
The elements of ${\mathcal F}_{min}$ define the centres of the first
round of blow-ups. In our case there is only one centre $L_{123}$.
The strict transforms of the elements of ${\mathcal F}$ which are
not in ${\mathcal F}_{min}$ define ${\mathcal F}_1$.
By abuse of notation we denote the strict transforms of $L_{ij}$ also
by $L_{ij}$:
\bq
 {\mathcal F}_1 & = & {\mathcal F} \backslash {\mathcal F}_{min} = \{ L_{12}, L_{23} \}.
\eq
The minimal elements of ${\mathcal F}_1$ define the centres of the
second round of blow-ups. In our case ${\mathcal F}_{1,min}={\mathcal F}_1$.
This procedure is continued by defining
${\mathcal F}_2 = {\mathcal F}_1 \backslash {\mathcal F}_{1,min}$ etc. until 
${\mathcal F}_j=\emptyset$.
In our case ${\mathcal F}_2 = \emptyset$ and we stop after the second
round of blow-ups.
Let us now investigate how the polynomial transforms under the blow-ups:
The first one blows up the centre $L_{123}$. In a coordinate patch
where $x_1>x_2$ and $x_1>x_3$ we have with $x_1=x_1'$, $x_2=x_1'x_2'$, $x_3=x_1'x_3'$:
\bq
 P & = & x_2^2 + x_1 x_3 = {x_1'}^2 \left( {x_2'}^2 + {x_3'} \right) = {x_1'}^2 P'.
\eq
We then have to blow up the centres of ${\mathcal F}_{1,min}$. The strict transform
of $L_{12}$ is not contained in the above coordinate patch, but the strict 
transform of $L_{23}$ is. Let us now consider what happens if we blow up
the centre given by the strict transform of $L_{23}$.
Looking in the coordinate patch with ${x'}_2 > {x'}_3$ we have with
${x_1'}={x_1''}$, ${x_2'}={x_2''}$, ${x_3'}={x_2''} {x_3''}$:
\bq
 P' & = & {x_2''} \left( {x_2''} + {x_3''} \right) = {x_2''} P''.
\eq
We see that $P''$ is not in monomialized form.
The proof of proposition 7.3 in \cite{Bloch:2005} relies crucially on the assumptions
(i) that only integrals of the form as in eq.~(\ref{localigusazetafct}) are considered,
(ii) that the graph under consideration is a graph with $n$ loops and $2n$ edges,
(iii) that all proper subgraphs are convergent.
In the present paper we do not make these assumptions.
In fact in our theorem 2 arbitrary polynomials may appear, which need not be related to Feynman graphs.
Our theorem is therefore a much stronger statement. 
In particular polynomials like eq.~(\ref{counterexample}) are allowed, which vanish along one direction
quadratically and along other directions linearly.

\end{appendix}

% ----------------------------------------------
% references
\bibliography{/home/stefanw/notes/biblio}
\bibliographystyle{/home/stefanw/latex-style/h-physrev3}

\end{document}